\documentclass[twocolumn,showpacs,preprintnumbers,amsmath,amssymb]{revtex4}

\usepackage{graphicx}
\usepackage{dcolumn} 
\usepackage{bm}      
\usepackage{color}

\usepackage{textcomp}

\begin{document}

\title{Contrast Mechanisms for the Detection of Ferroelectric Domains \\with
Scanning Force Microscopy}

\author{Tobias Jungk}
\email{jungk@uni-bonn.de} \affiliation{Institute of Physics,
University of Bonn, Wegelerstra\ss e 8, 53115 Bonn, Germany\\}
\author{\'{A}kos Hoffmann}
\affiliation{Institute of Physics, University of Bonn,
Wegelerstra\ss e 8, 53115 Bonn, Germany\\}
\author{Elisabeth Soergel}
\affiliation{Institute of Physics, University of Bonn,
Wegelerstra\ss e 8, 53115 Bonn, Germany\\}

\date{\today}

\begin{abstract}
We present a full analysis of the contrast mechanisms for the
detection of ferroelectric domains on all faces of bulk single
crystals using scanning force microscopy exemplified on hexagonally
poled lithium niobate. The domain contrast can be attributed to
three different mechanisms: i) the thickness change of the sample
due to an out-of-plane piezoelectric response (standard
piezoresponse force microscopy), ii) the lateral displacement of the
sample surface due to an in-plane piezoresponse, and iii) the
electrostatic tip-sample interaction at the domain boundaries caused
by surface charges on the crystallographic $y$- and $z$-faces. A
careful analysis of the movement of the cantilever with respect to
its orientation relative to the crystallographic axes of the sample
allows a clear attribution of the observed domain contrast to the
driving forces respectively.
\end{abstract}


\maketitle

\section{Introduction}

In the last decade, piezoresponse force microscopy (PFM) has become
a standard technique for imaging ferroelectric
domains~\cite{Alexe,Gru06,Kal06a}. This is mainly due to its ease of
use and nonetheless high lateral resolution of $< 20$\,nm without
any special sample preparation~\cite{Jun08a}. PFM is based on the
fact that ferroelectric materials are necessarily
piezoelectric~\cite{New}. Application of electric fields causes
thickness changes of the ferroelectric sample via the converse
piezoelectric effect. For PFM a scanning force microscope (SFM) is
operated in contact mode with a voltage applied to the conductive
tip. The thickness change induced by the electric field of the tip
is followed by the tip and leads to a deflection of the cantilever
that can be read-out as a vertical signal on the segmented
photo-detector. Because the thickness changes are in the order of a
few ten picometers an alternating voltage is applied to the tip and
the oscillations of the cantilever are read-out with a lock-in
amplifier. This contrast mechanism, based on the longitudinal
piezoelectric effect, is generally used in PFM measurements and will
be named "standard PFM" in the following.

In contrast to other visualization techniques~\cite{Soe05} PFM is
not restricted to specific crystallographic orientations. Recording
the torsion of the cantilever (a lateral signal on the segmented
photo-detector), thereby mapping an in-plane movement of the sample
surface, enables to record a domain structure also on crystal faces
not exhibiting an out-of-plane  response due to a  longitudinal
piezoelectric tensor element. An example is the visualization of the
domain structure on the non-polar face in periodically poled
KTiOPO$_4$ single crystals~\cite{Can03}.

Obviously, it is most appropriate to record the vertical and the
lateral signal simultaneously to obtain as much information as
possible on the domain structure of the sample. First experiments
using this detection scheme were carried out on bulk barium titanate
crystals~\cite{Abp98,Eng99}. Only a few studies on other bulk
crystals exhibiting a- and c-domains were reported  ever  since
(Bi$_4$Ti$_3$O$_{12}$~\cite{Yu05b,Kat07}, PbTiO$_3$~\cite{Oki01},
SrBi$_2$Ta$_2$O$_9$~\cite{Amo06}, PZN-PT~\cite{Abp02}, and
PMN-PT~\cite{Abp02,Zen04}). After all there are also a few
publications where simultaneous recording of the vertical and the
lateral signals on thin films is reported (BiFeO$_3$~\cite{Zav05},
LiNbO$_3$~\cite{Gau06}, PbTiO$_3$~\cite{Lop03,Zen05,Kal06b}, and
PZT~\cite{Gan02,Yu05a}).

In order to obtain full information on the domain configuration of a
crystal surface exhibiting $a$- and $c$-domains, however in general,
one single scan recording both read-out channels is not sufficient.
The vertical signal clearly distinguishes $+c$- and $-c$-domains.
The lateral signal depicts the in-plane $a$-domains. These, however,
can be oriented in four directions ($\leftarrow$ $|$ $\uparrow$ $|$
$\rightarrow$ $|$ $\downarrow$). Since within one image only two
in-plane orientations of the polarization ($\leftrightarrows$ or
$\uparrow\downarrow$) can be distinguished a second image of the
same area after rotation by $90^{\circ}$ is necessary to determine
the domain configuration completely. The crucial point for such
investigations is the relocation of the sample position after
rotation.   In case of  samples with  distinct topographic features
this is relatively easy and was performed on polycrystalline
particles of BiFeO$_3$~\cite{Shv07}, LaBGeO$_5$~\cite{Kal06b} or
thin film capacitors with top electrodes~\cite{Rod04}. But also for
samples with no specific topography, images at different angles were
recorded~\cite{Eng99,Amo06,Zav05,Zen05,Kal06b,Shv07,Rod04}.

The contrast in the lateral channel is generally attributed to a
shear movement of the sample caused by non-diagonal elements of the
piezoelectric tensor driven by the out-of-plane component of the
electric field $E_{\perp}$ from the tip
\cite{Abp98,Abp02,Gan02,Fel04}. Our measurements, however, strongly
support a different explanation: a sliding movement of the surface
induced by the in-plane components of the electric field
$E_{\parallel}$ together with an in-plane longitudinal piezoelectric
tensor element $d$. Although the electric field from the tip is
rotational symmetric the crystal deformations $\Delta t$ caused by
the opposed in-plane electric field components $E_{\parallel}$ do
not cancel out: $\Delta t = E\,t\,d$, where $t$ denotes the
thickness of the crystal and the direction of $E$ determines whether
$\Delta t$ is positive or negative, i.\,e.~resulting in an expansion
or a contraction of the crystal respectively.

In this contribution we present a thorough investigation of PFM
imaging on all crystal faces of hexagonally poled lithium niobate
(LiNbO$_3$) crystals. All faces show a distinct domain contrast,
which on the $x$- and $y$-faces strongly depends on the relative
orientation of the cantilever with respect to the crystallographic
axes. Three different contrast mechanisms were found to contribute
to the measured signals: (i) vertical piezoresponse of the sample
causing deflection of the cantilever (standard PFM), (ii) lateral
piezoresponse resulting in torsion and buckling on top of the domain
faces and (iii) electrostatic forces also leading to torsion and
buckling, however only at the domain boundaries~\cite{Jun06b}.  For
a clear attribution of the respective driving forces the angle
between the cantilever and the crystallographic axes must be known.
We therefore upgraded our PFM setup with a high precision
PC-controlled rotation stage. Performing angular-dependent
measurements of the vertical and the lateral signals -- both
recorded simultaneously -- on all three crystal faces we could
determine the surface displacements on each crystal face causing the
domain contrast.

This paper is organized as follows: Because the experimental
challenges are demanding we start with a detailed description of the
experimental methods (Sec.~\ref{sec:methods}). The next section
describes the experimental results (Sec.~\ref{sec:results}). Here we
firstly focus on our proposed model to explain a sliding of the
surface underneath the tip causing the domain contrast on the $x$-
and the $y$-faces (Sec.~\ref{sec:mov-sam}). Since this model has not
been described so far, but in contrary a different driving mechanism
is generally assumed to cause the surface deformation, this section
will be very detailed. In the following section (Sec.~\ref{sec:rot})
we present the experimental results to sustain the proposed surface
deformations by imaging all-faces of a two-dimensionally poled
lithium niobate crystal. The last section (Sec.~\ref{sec:y-face})
focusses once more of $y$-face imaging, showing most obviously the
proposed contrast mechanisms.

\section{Experimental methods}\label{sec:methods}

For the experiments we used a commercial scanning force microscope
(SMENA from NT-MDT) modified to allow the application of voltages to
the tip, and thus to be utilized as piezoresponse force microscope.
The instrument is operated in contact mode with an alternating
voltage applied to the tip ($U=5\,{\rm V_{rms}}, f=30-60$\,kHz), the
backside of the sample being grounded.

In Sec.~\ref{subsec:mov} we will describe the possibilities for
read-out of the cantilever movements. The excitation of the
movements depends on the orientation of the driving forces with
respect to the cantilever axis. We therefore upgraded our SFM setup
with a rotation stage described in the following
section~(Sec.~\ref{subsec:rot}). In addition we separately
investigated the friction between tip and
sample~(Sec.~\ref{subsec:fric}) and performed an accurate cross-talk
compensation~(Sec.~\ref{subsec:cross}). Finally, we briefly
summarize the properties of LiNbO$_3$ relevant for understanding the
origin of the domain contrast in SFM measurements, and in particular
present the sample we used for the
experiments~(Sec.~\ref{subsec:sam}).

\subsection{Movements of the cantilever}\label{subsec:mov}

\begin{figure}[ttt]
\includegraphics{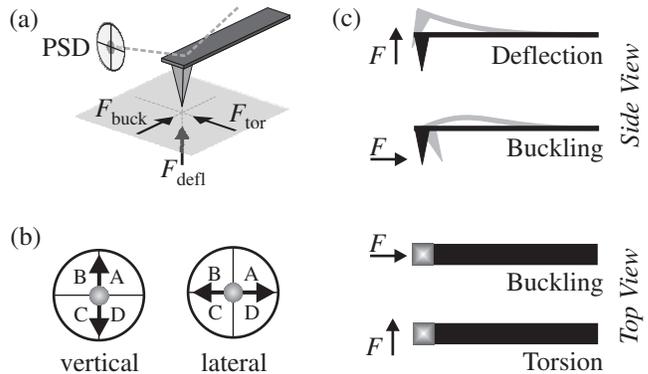}
\caption{\label{fig:Jungk1}
(a) Forces acting on the tip leading to deflection ($F_{\rm defl}$),
torsion ($F_{\rm tor}$), and buckling ($F_{\rm buck}$). (b)
Deflection and buckling are detected as vertical signals whereas
torsion results in a lateral signal. (c) The out-of-plane force
$F_{\rm defl}$ leads to a deflection of the cantilever which is
independent of its orientation with respect to the surface. For the
in-plane forces $F_{\rm tor}$ and $F_{\rm buck}$ the response of the
cantilever depends crucially on its orientation with respect to the
driving forces. PSD:~position sensitive detector.
}
\end{figure}

Basically the cantilever can perform three independent movements:
(i) deflection due to an out-of-plane force $F_{\perp}$ acting along
the axis of the tip, (ii) torsion caused by an in-plane force
$F_{\parallel}$ acting perpendicular to the cantilever axis and
(iii) buckling if an in-plane force $F_{\parallel}$ acts along the
cantilever axis (Fig.~\ref{fig:Jungk1}(a,c)). Deflection is
independent on the orientation of the cantilever with respect to the
sample. An in-plane force $F_{\parallel}$ acting on the tip results
in torsion $F_{\rm tor} = F_{\parallel} \sin \alpha$ and buckling
$F_{\rm buck} = F_{\parallel} \cos \alpha$ of the cantilever. Here
$\alpha$ denotes the angle between $F_{\parallel}$ and the
cantilever axis. Thus, a rotation of the sample by $90^{\circ}$
transforms a torsion signal into a buckling signal and vice versa.

Unfortunately a separate readout of all three signals is not
possible since both deflection and buckling are detected in the same
read-out channel as a vertical signal (Fig.~\ref{fig:Jungk1}b).
Torsion, however, can clearly be distinguished as it is read-out as
a lateral signal. In order to determine whether a vertical signal is
caused by deflection or by buckling, one has to perform measurements
at different angles $\alpha$. All angular dependent contributions to
the vertical signal can then be attributed to buckling movements. To
get the full picture on the piezoelectric responses of the sample we
recorded PFM data at different angles $\alpha$ with two lock-in
amplifiers, thus measuring vertical and lateral signals
simultaneously.

\subsection{Rotation sample stage}\label{subsec:rot}

\begin{figure}[ttt]
\includegraphics{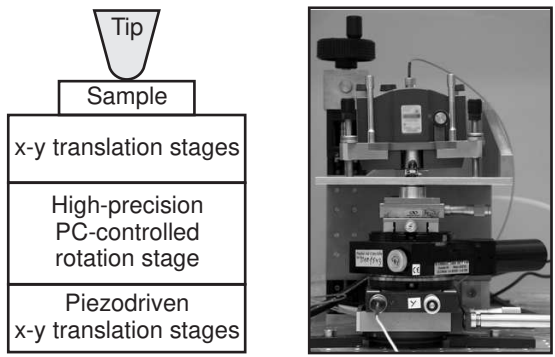}
\caption{\label{fig:Jungk2}
Schematics (left) and photo (right) of the setup for angular
dependent measurements. The piezo-driven translation stages at the
bottom allow a precise positioning of the rotation axis with respect
to the tip. The translation stages on top of the rotation stage are
used to position the sample.
}
\end{figure}

In order to determine the direction of the in-plane driving force
$F_{\parallel}$ acting on the tip, angular resolved measurements
were carried out. We therefore put the SFM on a home-built lifting
platform and mounted the sample on a high-precision, PC-controlled
rotation stage (Newport URM80). The latter was mounted on top of two
piezo-driven translation stages in order to position its rotation
axis precisely underneath the tip (Fig.~\ref{fig:Jungk2}). The
obtained accuracy was such that on a full revolution the tip
describes a circle of $<3$\,\textmu m radius on the surface. Another
set of translation stages, mounted on top of the rotation stage, was
used to position the sample underneath the tip. The whole setup was
found to be stable enough to operate the SFM while rotating the
sample.

In this setup the adjustment of the pivot of the rotation stage with
respect to the  position of the  tip is very crucial. It has to be
repeated for every new cantilever since its mounting is not precise
to an accuracy of a few microns. For  a  coarse alignment  the
displacement of a 20\,\textmu m wide cross was observed  during
rotation with the help of an optical microscope. By repeated
adjustment of the pivot after  $180^{\circ}$ rotations an accuracy
of approx.~20\,\textmu m was achieved. The fine alignment was
conducted with a periodically poled LiNbO$_3$ (PPLN) sample with a
domain width of 4\,\textmu m. Comparing two PFM images recorded
before and after rotating the sample by $180^{\circ}$ allowed to
determine the misalignment and thereafter to correct it with the
piezo-driven translation stages underneath the rotation stage
(Fig.~\ref{fig:Jungk2}).

This setup allows to record SFM images at specific angles between
the cantilever and the crystallographic axes of the sample.
Furthermore, we could obtain angular dependent measurements with the
tip stationary, named "rotation scans" in the following. They were
carried out with the help of a LabView programm addressing the
rotation stage as well as the lock-in amplifiers. For data
acquisition we waited $100 - 1000$\,ms at every new angle ($10
\times$ the time constant of the lock-in amplifier) to let the
microscope calm down from the braking of the rotation stage.

\subsection{Friction between tip and sample}\label{subsec:fric}

\begin{figure}[ttt]
\includegraphics{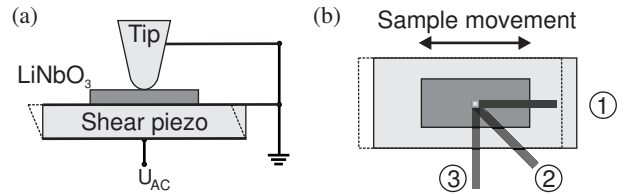}
\caption{\label{fig:Jungk3}
(a) Test-setup for determining the friction between tip and  $\rm
LiNbO_3$ crystal surface. (b) Three selected orientations of the
cantilever. In \textcircled{1} only bucking can be observed, in
\textcircled{3} the cantilever is oriented such that torsion occurs
and \textcircled{2} shows the situation for exciting both movements.
}
\end{figure}

Because in-plane forces act on the tip, special attention has to be
payed to the friction between tip and sample surface. Therefore we
realized a test setup depicted in Fig.~\ref{fig:Jungk3}(a) where we
mounted a single domain LiNbO$_3$ crystal on top of a shear piezo
driven with an alternating voltage $U_{\rm AC}$. In order to avoid
any electrostatic fores between tip and sample, both were grounded.
In this configuration, only in-plane forces act on the tip,
resulting in torsion and/or buckling of the cantilever, depending on
the relative orientation between the sample movement and the
cantilever axis (Fig.~\ref{fig:Jungk3}(b)). The tip was found to
follow the surface movement perfectly for an amplitude $A < 300$\,pm
and a frequency $f<100$\,kHz.

\subsection{Cross-talk compensation}\label{subsec:cross}

The aim to accurately attribute the measured signals to the
respective driving forces requires the cross-talk-free detection of
the vertical and the lateral signals. Usually, however, due to a
misalignment between the plane of the read-out laser beam and the
orientation of the position sensitive detector, the two signals are
not fully separated. The setup with the shear piezo can be used to
adjust for the cross-talk with the tip being in contact with the
sample. Torsion and buckling can be excited separately through the
shear piezo movement along with the angular adjustment using the
rotation stage. The false signals caused by cross-talk are then
electronically suppressed~\cite{Hof07}. This way of cross-talk
compensation is advantageous  insofar, as it is performed under the
same conditions  as  the measurements carried out later on.

\subsection{Sample and sample preparation}\label{subsec:sam}

\begin{figure}[ttt]
\includegraphics{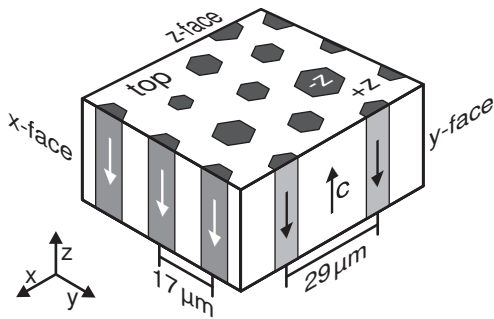}
\caption{\label{fig:Jungk4}
Schematics of the hexagonally poled $\rm LiNbO_3$ sample (HexLN)
used for the investigations. All faces are optically polished. The
periodicity of the corresponding PPLN structures on the side faces
are indicated together with the orientation of the polar $c$-axis.
}
\end{figure}

The experiments were carried out with congruently melting lithium
niobate crystals ($\rm LiNbO_{3}$). Here, we briefly summarize the
crystals main properties: The surface polarization charge on the
$z$-face is $P_s=0.7$\,C/m$^2$~\cite{Landolt}. This value being very
high, a compensation of the charges takes place, which is
assumed~\cite{Lik01} and experimentally verified~\cite{Jun06b} to be
of the order of a factor of 1000. This value strongly depends on the
exact crystal composition as well as the environmental conditions,
i.\,e.~mainly humidity. We also have evidence for surface charging
on the $y$-faces which is of the order of $\frac{1}{3}$ to $
\frac{1}{4} P_s$. Although this has not been explicitly reported so
far, however, it has already been observed~\cite{Gou02}  and it is
also very possible from theoretical considerations~\cite{Tas79}. We
claim the presence of a surface charging on the $y$-face because of
the observation of long-range ($>10$\,\textmu m) forces  during
tip-sample approach. The latter was found to behave similar than on
the $z$-face but completely different from non-charged surfaces like
glass or the $x$-face of LiNbO$_3$. Furthermore the contrast
observed at the domain boundaries on the $y$-faces
(Secs~\ref{sec:rot},\ref{sec:y-face}) can only be explained by a
domain specific surface charging~\cite{Jun06b}.

Because of being ferroelectric, LiNbO$_3$ is also piezoelectric and
the piezoelectric coefficients can be found in the
literature~\cite{Landolt}: $d_{22} = 21$\, pm/V, $d_{33} = (6 -
21)$\, pm/V, $d_{15} = 69$\, pm/V and $d_{31} = - 1$\, pm/V. We
would like to point out three facts:

\begin{itemize}
    \item[-] There is no longitudinal piezoelectric tensor
    element along the $x$-axis ($d_{11} = 0$).
    \item[-] For the longitudinal
    piezoelectric tensor element $d_{33}$ along the $z$-axis, the
    reported values vary considerably (by a factor of three).
    \item[-] The values given above apply for bulk crystals in a homogeneous electrical field,
    i.\,e.~samples covered with large top electrodes. In the case of
    PFM, however, the tip acts as top electrode, leading to a strongly inhomogeneous electric
    field expanding only a few microns into the crystal. As a
    consequence, clamping reduces the electromechanical
    response of the crystal. Since clamping depends on elasticity, this effect is not the same for all
    crystal orientations. We measured the ratio between the longitudinal piezoelectric
    coefficients as they are determined by PFM using single domain $y$- and $z$-cut samples to be $d_{22}:d_{33} = 1:5$.
\end{itemize}

For the experiments we used a two-dimensionally periodically poled
lithium niobate ($\rm LiNbO_3$) crystal as depicted in
Fig.~\ref{fig:Jungk4}. All faces were optically polished. The
different period lengths of the domain structures on the $x$- and on
the $y$-faces allow a unambiguous identification of the face under
investigation. The sample size was $(5 \times 7 \times
0.5)$\,mm$^3$.

To exclude any effect from a possible mis-orientation of the sample
cuts relative to the crystallographic axes, we also prepared samples
were half of the $x$- and $y$-faces were polished at angles of
5$^{\circ}$ and 10$^{\circ}$. However, the slants did not affect the
results.

During the experiments it turned out that special care has to be
taken of the mounting of the sample. The lock-in detection technique
being extremely sensitive, any inhomogeneity of the grounding
electrode was observed to have an influence on the measured signals,
presuming an angular dependence when rotating the sample. For the
measurements on the $z$-face, the crystal was simply led on a large
metal plate (2\,cm diameter). For investigating the side faces we
mounted the sample with the help of two additional plastic brackets.

\section{Experimental results}\label{sec:results}

In order to avoid a misunderstanding in the following we would like
to point out some basics here. Applying a voltage $U$ to a
piezoelectric sample of thickness $t$ leads to an electric field $E
= U/t$ inside that sample causing a piezoelectric deformation
$\Delta t = E\,d\,t = U\,d$ with $d$ being the appropriate
piezoelectric tensor element. Depending on the direction of $E$ the
sample experiences contraction ($\Delta t < 0$) or expansion
($\Delta t > 0$). The same applies also in case of an inhomogeneous
electric field, like the one generated by the tip of the PFM, since
$\int_0^t E \rm{d} s = U$~\cite{Jun07c}.

\subsection{Movements of the sample surface}\label{sec:mov-sam}

\begin{figure}[ttt]
\includegraphics{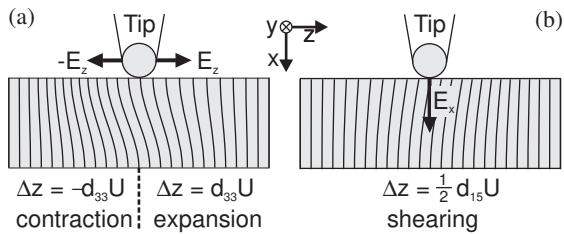}
\caption{\label{fig:Jungk5}
Possible origins for the lateral movement of the $x$-surface of a
single domain LiNbO$_3$ sample underneath the tip (applied voltage
$U$). In (a) the deformation of the crystal is due to a contraction
and an expansion caused by the electrical field components $\pm
E_z$. (b) Shearing caused by the electric field component $E_x$ is
strongly suppressed by clamping.
}
\end{figure}

Applying a voltage $U$ to the tip results in a first approximation
in a radial field distribution when approximating the tip by a
sphere. Usually in PFM, only the out-of-plane component $E_{\perp}$
of the electric field, perpendicular to the sample surface and thus
along the axis of the tip is taken into account, causing a thickness
change of the sample. A few publications also discuss shear
deformations of the sample caused by
$E_{\perp}$~\cite{Abp98,Abp02,Gan02,Fel04,Kal06b}. All other
components of the electric field $E_{\parallel}$ in plane of the
sample surface are generally presumed to cancel out due to
rotational symmetric considerations. This argument, however, does
not hold. In contrary, the contributions of the piezoelectric
deformations caused by opposed electric field components $\pm
E_{\parallel}$ add up to a net movement of the surface. Thus for a
full analysis of the deformation of the sample surface all
components of the electric field have to be taken into account. For
our analysis we consider exemplarily the 5~principal directions of
the electric field choosing the notation such that they follow the
crystallographic axes. I.\,e.~when the tip is on top of a $x$-face,
the electric field component $E_{\perp}$ is called $E_x$. The in
plane electric field components $E_{\parallel}$ are consistently
called $\pm E_y$ and $\pm E_z$.

A step-by-step analysis of the deformation of the $x$-face can be
accomplished as follows (Fig.~\ref{fig:Jungk5}):
\begin{itemize}
\item[-] The effect of $E_x$ on the piezoresponse of the sample
due to the longitudinal tensor element $d_{11}$ can be easily
analyzed since for LiNbO$_3$ $d_{11}=0$, thus the thickness change
is $\Delta x = d_{11}\,U=0$.
\item[-] Whereas the electric field component $+E_z$ causes expansion ($\Delta z =
d_{33}\,U$) the opposed component $-E_z$ leads to a contraction of
the crystal ($\Delta z = -d_{33}\,U$). As a result a net sliding of
the surface along the $z$-axis occurs (Fig.~\ref{fig:Jungk5}(a)).
\item[-] In an analogous way, the electric field components $\pm E_y$ excite the longitudinal tensor
element $d_{22}$ which leads to a sliding of the surface along the
$y$-axis.
\item[-] The effect of $E_x$ on the piezoresponse of the sample
due to the shear tensor element $d_{15}$ is schematically depicted
in Fig.~\ref{fig:Jungk5}(b). Such a deformation, however, is
strongly suppressed. This can be understood as follows: Within the
whole volume experiencing the electric field $E_x$ the crystals
deformation is of same type, e.\,g.\,a shear to the right. Since
this volume is restricted compression and extraction occurs at its
limits leading to clamping. The fact that the electric field $E_x$
decays with $1/r^2$ ($r$: tip radius) and therefore the limits are
not rigid does not play a role. The only difference is that clamping
is dispersed along the decay length of the electric field $E_x$, the
overall effect, however, persists.
Note that this situation is very different from the one in
Fig.~\ref{fig:Jungk5}(a), where one side wants to contract and the
other side wants to expand.
\end{itemize}

\begin{figure}[ttt]
\includegraphics{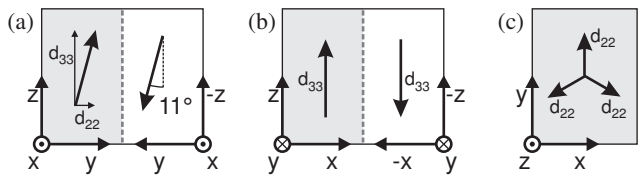}
\caption{\label{fig:Jungk6}
In-plane displacement of the different surfaces of a LiNbO$_3$
crystal, shown for a bi-domain crystal. (a) On the $x$-face, the
contributions from $d_{22}$ and $d_{33}$ add up vectorially. (b) On
the $y$-face, only the $d_{33}$ tensor component leads to a sliding
of the surface. (c) The movement on the $z$-face caused by $d_{22}$
reflects the trigonal symmetry of the material.
}
\end{figure}

As a result of the considerations listed above, the $x$-face
performs a sliding movement underneath the tip at an angle of
$\theta=\arctan\left(d_{22}/d_{33}\right) = 11^{\circ}$ relative to
the crystallographic $z$-axis, inserting the numbers for the
piezoelectric tensor elements determined previously
(Fig.~\ref{fig:Jungk6}(a)). A contribution of shearing to the
surface displacement would lead to another angle. Its precise
measurement is thus the key experiment to sustain the proposed model
of expansion \& contraction and suppressed shearing.

Similar considerations apply for the $y$-face
(Fig.~\ref{fig:Jungk6}(b)). Note that there is a net thickness
change of the sample $\Delta y = d_{22}\,U$. The $E_{\parallel}$
components of the electric field, lead to $\Delta x = d_{11}\, U =
0$ since $d_{11}=0$ and $\Delta z = d_{33}\, U$ resulting in a net
sliding of the surface along the $z$-axis.

For the $z$-face, there is a net thickness change $\Delta z =
d_{33}\,U$, yielding the standard PFM signal. Upon closer
inspection, we observed an extremely small piezomechanical in-plane
response, reflecting the trifold symmetry of the crystals $y$-axes.
This movement of the surface can be attributed to the fact that the
tip has an inclination with respect to the surface of approx.\
$20^{\circ}$. This leads to a slightly asymmetric electric field
distribution and thus the contributions from the three identical
$y$-axes of the crystal do not cancel each other. In
Fig.~\ref{fig:Jungk6}(c) the in-plane displacements on the three
crystal faces of LiNbO$_3$ are shown.

\subsection{Rotation scans}\label{sec:rot}

\begin{figure*}[ttt]
\includegraphics{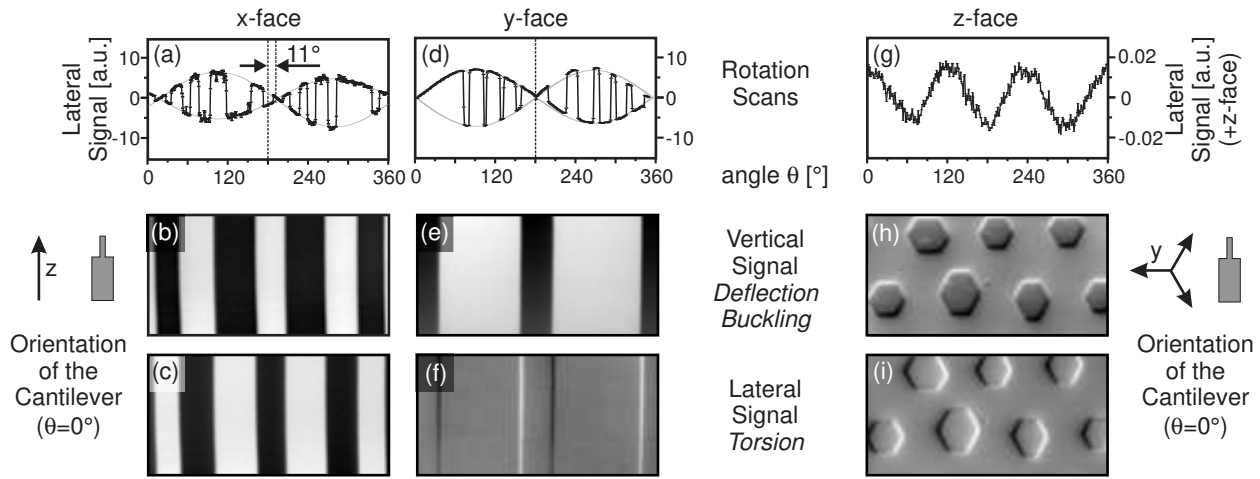}
\caption{\label{fig:Jungk7}
Top row:~Lateral signal during $360^{\circ}$\,rotation scans on all
three faces of the HexLN sample. Center row:~vertical image and
bottom row:~lateral image of the three faces respectively. Left
column: $x$-face, middle column: $y$-face and right column:
$z$-face. For the $x$- and the $y$-faces, the angle is defined
between the cantilever C and the $z$-axis of the crystal
($\theta=0^{\circ}: {\rm C}\parallel z$). For the $z$-face an angle
of $0^{\circ}$~corresponds to a cantilever being orientated
perpendicular to one of the three $y$-axes ($\theta=0^{\circ}: {\rm
C}\perp y$) Image size for all six images: 60\,\textmu m $\times$
30\,\textmu m.
}
\end{figure*}

Figure~\ref{fig:Jungk7} summarizes our experimental results. We
always recorded the vertical and the lateral signal simultaneously,
thus (b and c), (e and f), and (h and i) were recorded during the
same scan. Obviously, all images show a clear domain contrast,
whereas (b,c, and e) show only the domain faces, (f and i) only the
domain boundaries, and (h) a combination of both. On every face, we
also performed rotation scans, i.\,e., we rotated the sample by
$360^{\circ}$ during data acquisition. To clearly attribute the
recorded signals to in-plane driving forces, we show the data for
the lateral signal (a,d,g). For the $x$- and the $y$-face, the pivot
of the rotation axis was set such that the tip performed a circle of
several 10\,\textmu m radius on top of the surface. Thereby, we
could record the data for both domain faces simultaneously, seen as
jumps in the graphs. The scan on the $z$-face was performed on a
single $+z$ domain face.

For understanding the content of Fig.~\ref{fig:Jungk7} we start with
the last column depicting the measurements on the $z$-face.
Fig.~\ref{fig:Jungk7}(h) shows the standard PFM deflection signal,
with the $-z$ faces of the hexagonal domains appearing dark on the
bright $+z$ face. The white borders on the upper edges and the dark
ones a the lower edges are due to a buckling movement of the
cantilever at the domain boundaries. This is caused by the
electrostatic interaction between the charged tip and the
electrostatic field at the domain walls. The latter is due to the
differently orientated domains carrying surface polarization charges
of opposite polarity. In the torsion channel recorded in
Fig~\ref{fig:Jungk7}(i) only the domain boundaries are seen. A
detailed description of the lateral electrostatic force microscopy
can be found elsewhere~\cite{Jun06a}.

The rotation scan on the $z$-face (Fig.~\ref{fig:Jungk7}(g)) shows a
very small, but clearly resolvable lateral signal. As expected from
the considerations described above, the three maxima reflect the
trigonal symmetry of LiNbO$_3$ indicating the three identical
crystallographic $y$-axes.

Investigations of the $y$-face of the crystal yield the following
results: Upon perfect alignment of the cantilever parallel to the
crystallographic $z$-axis, the lateral signal shows only the domain
boundaries (Fig.~\ref{fig:Jungk7}(f)) due to the electrostatic
interactions~\cite{Jun06a}. Rotating the sample, however, reveals a
lateral movement on top of the domain faces, with a maximum at an
angle of $90^{\circ}$ (Fig.~\ref{fig:Jungk7}(d)). This can be
attributed to the sliding of the surface caused by the longitudinal
deformation along the $z$-axis. Torsion and buckling, however, being
caused by the same driving force, their maximum amplitudes occur at
an relative angle of $90^{\circ}$. Thus, the contrast in
Fig.~\ref{fig:Jungk7}(e) is due to a superposition of (i) a
deflection of the cantilever caused by the standard PFM ($\Delta y =
d_{22}\,U$) and (ii) its buckling caused by a sliding along the
$z$-axes of the crystal.

Finally, figure~\ref{fig:Jungk7}(b) shows the most surprising
result: a clear domain contrast on the $x$-face of LiNbO$_3$.
Although unexpected at first sight (because $d_{11}=0$), this is
consistent with the considerations exposed above. According to our
expansion-contraction-model sliding of the surface should occur at
an angle of $\theta = 11^{\circ}$. For images taken at $\theta =
0^{\circ}$ the cantilever performs both a buckling and a torsion
movement as seen in the images in Figs.~\ref{fig:Jungk7}(b and c).
The exact angle of the maximum surface displacement could be
confirmed by a rotation scan shown in Fig.~\ref{fig:Jungk7}(a) where
we found a shift of the zero-crossing of $11^{\circ}$. Note the
adjustment of the cantilever axis with respect to the $z$-axis is
performed by recording large PFM images and is precise to about
$1^{\circ}$.

\subsection{Lateral y-face imaging}\label{sec:y-face}

\begin{figure}[ttt]
\includegraphics{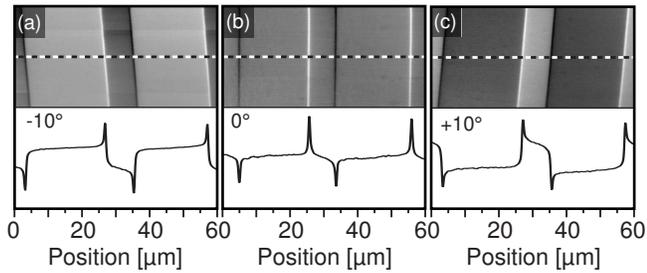}
\caption{\label{fig:Jungk8}
Lateral signal on a $y$-face recorded for three different
orientations between cantilever and $z$-axis: a) $-10^{\circ}$, b)
$0^{\circ}$, and c) $+10^{\circ}$.
}
\end{figure}

Finally, a clear demonstration for the contribution of the two
different contrast mechanisms to the lateral signal (surface
displacement and electrostatic interaction) on a $y$-face can be
seen in Fig.~\ref{fig:Jungk8}. We therefore recorded three images at
different angles between the cantilever axis and the $z$-axis of the
crystal. With the cantilever oriented parallel to the $z$-axis of
the crystal only the electrostatic signal at the domain boundaries
can be observed (Fig.~\ref{fig:Jungk8}(b)). The line scan 
shows no contrast of the domain faces. When the sample is rotated by
$-10^{\circ}$ with respect to the cantilever axis also the domain
faces show a contrast due to a contribution of the $z$-directed
sliding of the surface. Note that the wide domains appear brighter
than the narrow ones (Fig.~\ref{fig:Jungk8}(a)). If the sample is
rotated by $+10^{\circ}$ the projection of the $z$-directed movement
to the torsion of the cantilever is inverse, thus the wide domains
show up as dark stripes while the narrow ones are bright
(Fig.~\ref{fig:Jungk8}(c)).

\section{Conclusions}

In this contribution, we have carried out a thorough investigation
of the origin of the domain contrast on all faces of of LiNbO$_3$
single crystals. A clear domain contrast could be observed on all
crystal faces, in the vertical as well as in the lateral read-out
channel. Upon close inspection, the domain contrast could be
attributed to a superposition of three driving forces: (i) standard
piezoresponse force microscopy (PFM) originating from a thickness
change of the sample, (ii) an  piezomechanically caused lateral
displacement of the surface due to an
expanding-contraction-movement, and (iii) the electrostatic
interaction of the charged tip with the electric fields on top of
the crystals $y$- and $z$-faces.

In order to perform these experiments, we upgraded our scanning
force microscope (SFM) mounting the samples on a high-precision
rotation stage. It has thus become possible to investigate a well
defined position on top of  the sample surface at different
orientations of the cantilever with respect to the sample. This
technical issue in combination with the complete understanding of
the detection mechanism now allows to determine unambiguously the
orientation of the polar axis any arbitrarily orientated
ferroelectric particle and for specifically cut piezoelectric
samples to determine their symmetry axis.

\section{Acknowledgments}
We thank C.~Gawith from the Optoelectronics Research Centre,
University of Southampton (UK) for providing the HexLN sample.
Financial support of the Deutsche Telekom AG is gratefully
acknowledged.

\end{document}